\documentclass[aps,pre,
showpacs,floatfix]{revtex4}
\usepackage{amssymb,amsmath,amsfonts,graphicx}
\newcommand{\be}{\begin{equation}}
\newcommand{\ee}{\end{equation}}
\newcommand{\bea}{\begin{eqnarray}}
\newcommand{\eea}{\end{eqnarray}}

\begin{document}

\title{Freezing of Nonlinear Bloch Oscillations in the Generalized Discrete
Nonlinear Schr\"{o}dinger Equation}
\author{ {\bf F. J. Cao$^{1,2}$}}
\affiliation{
$^{1}$ Centro de F\'{\i}sica Te\'orica e Computacional,\\
Universidade de Lisboa, Complexo Interdisciplinar,\\
Av. Prof. Gama Pinto 2, Lisboa 1649-003, Portugal \\ \\
$^{2}$ Grupo Interdisciplinar de Sistemas Complejos (GISC) and \\
Departamento de F\'{\i}sica At\'omica, Molecular y Nuclear.\\ 
Universidad Complutense de Madrid. \\ E-28040 Madrid.
Spain.}

\date{June 4, 2004}

\begin{abstract}
 The dynamics in a nonlinear Schr\"odinger chain in an {\em
homogeneous} electric field is studied. We show that {\em discrete
translational invariant} integrability-breaking terms can freeze
the Bloch nonlinear oscillations and introduce new faster
frequencies in their dynamics. These phenomena are studied by
direct numerical integration and through an adiabatic
approximation. The adiabatic approximation allows a description in
terms of an effective potential that greatly clarifies the
phenomena.
\end{abstract}


\maketitle

\section{Introduction}

The study of the generalized discrete nonlinear Schr\"{o}dinger (GDNLS)
equation, introduced by Salerno \cite{salerno} as a model
providing one-parametric transition between discrete nonlinear
Schr\"{o}dinger (DNLS) equation and exactly integrable
Ablowitz-Ladik (AL) model \cite{AL}, is attracting increasing
interest, due to the relevance of lattice dynamics in various
fields of physics, as condensed matter, fiber optics physics,
molecular biology (see e.g. \cite{Scott} and references therein)
and recently Bose-Einstein condensate (see e.g \cite{AKKS} and
references therein).

One of the most interesting phenomena, which can be observed in
the different discretizations of the nonlinear Schr\"{o}dinger
equation is the so-called Bloch oscillations, that appear when a
linear force is applied to a solitary wave solution. Such
oscillations in the integrable AL model with a linear force have
been discovered in~\cite{BLR}, numerically found in the
DNLS~\cite{Scharf}, and interpreted as Bloch oscillations, using
the analogy with the solid state physics, in \cite{KCV}. Later on,
Bloch oscillations were studied in the GDNLS equation
\cite{CaiElectric} and in the presence of
impurities~\cite{KonoAdiabatic}. Bloch oscillations have been
observed experimentally in an array of waveguides~\cite{array},
found also to exist in the case of a dark soliton~\cite{kondark}
and in a totally discretized (i.e. discrete-time discrete space)
nonlinear Schr\"{o}dinger equation~\cite{kondisc}

In the present article we report some peculiarities of the Bloch
oscillations of bright solitons in the GDNLS equation. More
specifically we show that the model preserves the Bloch
oscillations, the amplitude of which however displays in certain
cases an abrupt change when the parameter governing transition
between AL and DNLS models is changed.

Section \ref{sDNLS} presents the model, {\em i.e.}, the GDNLS
equation with discrete translational invariant
integrability-breaking terms in an external homogenous electric
field, and the soliton initial conditions considered for this
model. In Section \ref{sAdiabatic} we present the adiabatic
approximation, that will prove to be very useful to understand the
dynamics. It allows a description in terms of an effective
potential. In Section \ref{sDynamics} we describe the main
ingredients of the dynamics: the Bloch oscillations and the
freezing of the Bloch oscillations, and provide an explanation of
this features in terms of the adiabatic effective potential.
Finally, in the conclusions (Section
\ref{sConclusions}), we summarize  the main results and discuss
their relevance.

\section{The model} \label{sDNLS}

The equation of motion for the system we are dealing with reads
\be
i \frac{d\psi_{n}}{dt} + (1+|\psi_{n}|^2)(\psi_{n+1}-\psi_{n-1}) -
2 \chi n \psi_{n} -
\nu (\psi_{n+1}+\psi_{n-1}-2\psi_{n})|\psi_{n}|^2  = 0 \;.
\label{DNLSeqf}
\ee
where $\nu$ is the integrability-breaking
parameter providing one-parametric transition between the AL
model~\cite{AL} ($\nu=0$) and  the DNLS model ($\nu=1$), and
$\chi$ is a parameter defining the strength of the linear force.
In particular, at $\nu = 0$ Eq. (\ref{DNLSeqf}) is integrable and
has the exact bright soliton solution~\cite{KCV}
\be
\psi_{n}(t) =
\frac{\sinh(2\,w)\;\exp\left\{i\left[{\frac {\cosh(2\,w)\sin(2\,\chi\,t)}{\chi}}-2\,
\chi\,nt\right]\right\}}{\cosh\left\{2\,w(n-\xi_0)-{\frac {
\sinh(2\,w)\left[\cos(2\,\chi\,t)-1\right]}{\chi}}\right\} }
\ee
where $w$ and $\xi_0$ are constant parameters of the solutions
($\xi_0$ can be interpreted as the initial position of the soliton
center, and $1/w$ as the soliton width).

Our purpose is to study how the $\nu$-term changes the soliton evolution;
{\em i.e.}, we consider the evolution for $\nu \neq 0$ of initial
conditions of the form
\be
\psi_{n}(0) = {\frac {\sinh(2\,w)}{\cosh[2\,w (n-\xi_0)]}}
\ee

\section{Adiabatic approximation} \label{sAdiabatic}

\begin{figure}[h]
\[
\includegraphics[width=6cm]{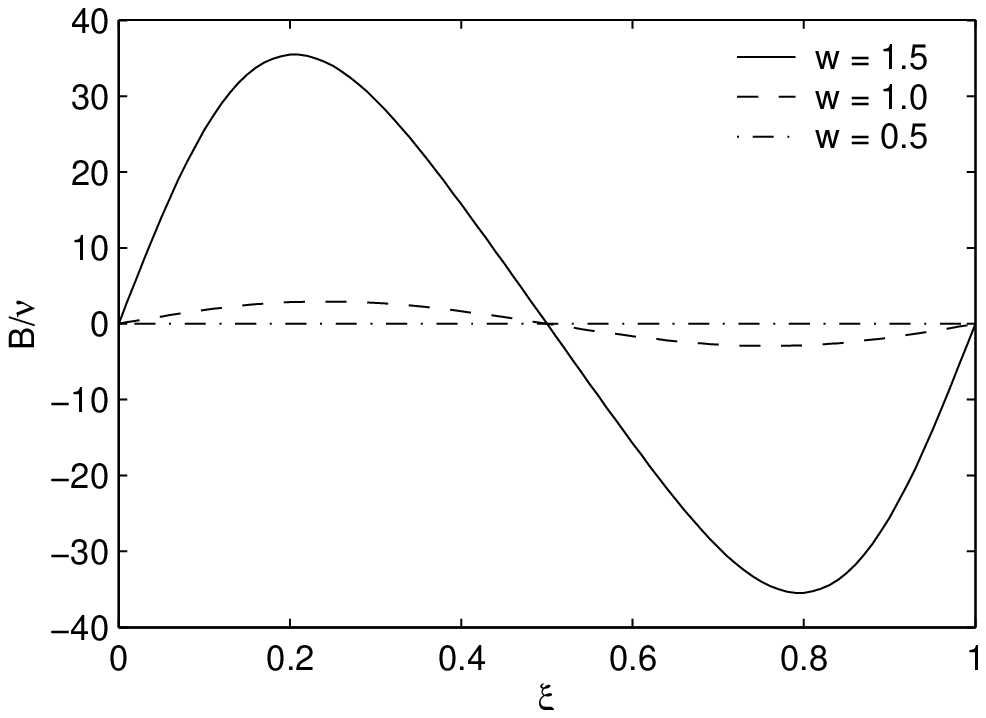}
\includegraphics[width=6cm]{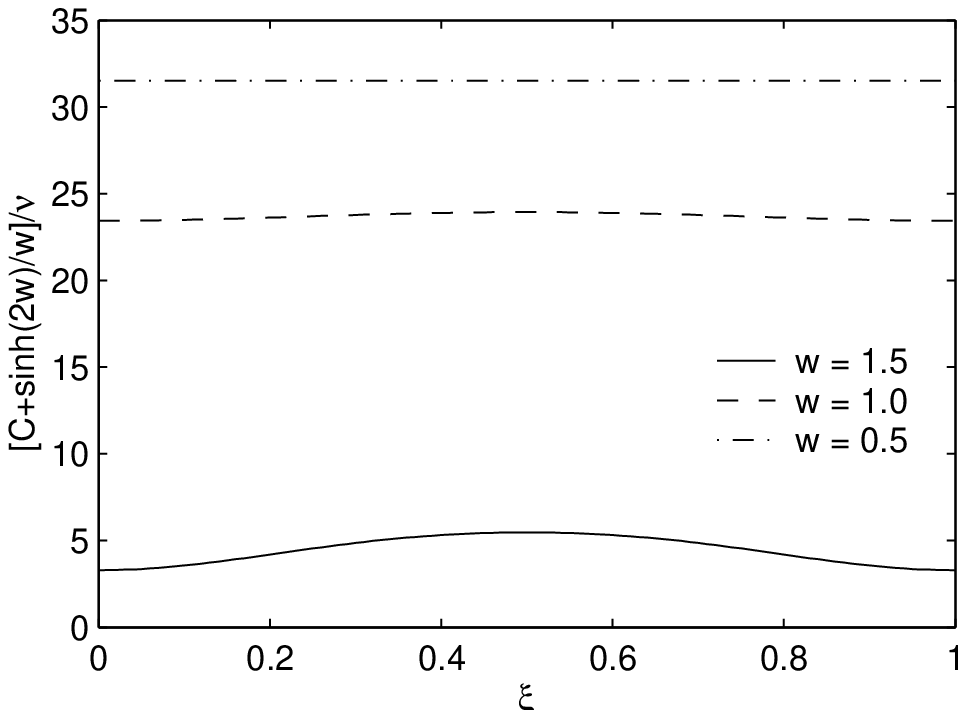}
\]
\caption{{\em (a)} $B /\nu$ and {\em (b)} $ [C+\sinh(2w)/w]/\nu $
as a function of $ \xi $ for various values of $w$. }
\label{fB}
\end{figure}

Let us start with the analysis of the problem when it is close to
integrable, {\em i.e.}, when $|\nu|\ll 1$. We employ the
perturbation theory (in the presence of a linear force it was
developed in~\cite{KonoAdiabatic}) and limit ourselves to the
adiabatic approximation. This means that the term $\nu
(\psi_{n+1}+\psi_{n-1}-2\psi_{n})|\psi_{n}|^2$ is considered as a
perturbation of the AL model. As we have previously seen, for $
\nu = 0 $ there are exact solutions of the form
\be \label{AdiabaticForm}
\psi_n = -i \frac{ \sinh(2w)\, e^{-2i\theta(n-\xi)+i\varphi}\,
e^{-i(2n+1)\chi t} }{ \cosh[2w(n-\xi)] }\;.
\ee
In the adiabatic approximation we compute the time evolution of
the parameters $w$, $\xi$, $\theta$ and $\varphi$; while keeping
the functional form of Eq.~\eqref{AdiabaticForm} fixed.
The equation of motion \eqref{DNLSeqf} can be written as
\be
i \frac{d\psi_{n}}{dt} + (1+|\psi_{n}|^2)(\psi_{n+1}-\psi_{n-1}) -
2 \chi n \psi_{n} =
\nu (\psi_{n+1}+\psi_{n-1}-2\psi_{n})|\psi_{n}|^2 \;.
\label{DNLSeqfi}
\ee
with the $\nu$-term considered as a perturbation. This approach
gives the following evolution equations for the parameters
\bea
\frac{d\theta}{dt} &=& A(w,\xi)\, \cos[2(\chi t + \theta)] + B(w,\xi) \label{dtheta} \\
\frac{d\xi}{dt} &=& C(w,\xi)\, \sin[2(\chi t + \theta)] \label{dxi} \\
\frac{dw}{dt} &=& D(w,\xi)\, \sin[2(\chi t + \theta)] \label{dw}
\eea
where
\bea
A(w,\xi) &=& \frac{\nu}{2}\,\sinh^4(2w)
\cr &&\times \sum _{n=-\infty }^{\infty}
  \frac {\sinh[2w(n-\xi)]}{\cosh[2w(n+1-\xi)] \cosh^2[2w(n-\xi)]
  \cosh[2w(n-1-\xi)]}
\cr &&\quad\quad\quad\quad \times
 \left\{\frac1{\cosh[2w(n+1-\xi)]}+\frac1{\cosh[2w(n-1-\xi)]}
  \right\}
\\
B(w,\xi) &=& -\nu\,\sinh^4(2\,w)
\cr &&\times \sum _{n=-\infty }^{\infty }{
  \frac {\sinh[2w(n-\xi)]}{\cosh[2w(n+1-\xi)]
  \cosh^3[2w(n-\xi)] \cosh[2w(n-1-\xi)]}}  \label{B}
\\
C(w,\xi) &=& -\frac{\sinh(2\,w)}{w}
  \left(1+\frac{\nu}2\,\sinh^3(2\,w)
 \right. \cr &&\times \left.
  \sum _{n= -\infty}^{\infty } \frac{(n-\xi)}
  {\cosh[2w(n+1-\xi)]\cosh[2w(n-\xi)]\cosh[2w(n-1-\xi)]}
  \right.
  \cr &&\quad\quad\quad\quad \times \left. \left\{\frac1{\cosh[2w(n+1-\xi)]}
  -\frac1{\cosh[2w(n-1-\xi)]} \right\} \right)  \label{C}
\\
D(w,\xi) &=& -\frac 12\,\nu\,\sinh^2(2\,w)
\cr &&\times \sum _{n=-\infty }^{\infty }
  \frac1{\cosh[2w(n+1-\xi)]\cosh[2w(n-\xi)]\cosh[2w(n-1-\xi)]}
\cr &&\quad\quad\quad\quad \times
  \left\{\frac1{\cosh[2w(n+1-\xi)]}-\frac1{\cosh[2w(n-1-\xi)]}\right\} \;.
\eea
From these expressions one can see that the coefficients $A$, $B$,
$C$ and $D$ are periodic in $\xi$ with period $1$. The dynamics of
$\theta$, $\xi$ and $w$ is decoupled from the evolution of the
phase $\varphi$, and $\varphi$ is merely slaved to the dynamics of
the previous parameters. We have seen numerically that for a wide
range of parameters $A(w,\xi) = 0$ and $D(w,\xi) = 0$ (for $ \xi $
integer of halfinteger it is easy to prove it analytically using
that the terms of the series are odd-functions in $n$),
therefore the adiabatic equations of motion reduce to
\bea
\frac{d\theta}{dt} &=& B(w,\xi) \label{dthetar} \\
\frac{d\xi}{dt} &=& C(w,\xi)\, \sin[2(\chi t + \theta)] \label{dxir} \\
\frac{dw}{dt} &=& 0 \;. \label{dwr}
\eea
The corrections in $B$ and $C$ due to the $\nu$-term, $ B / \nu $
and $ [C+\sinh(2w)/w]/\nu $ [Eqs.~\eqref{B}-\eqref{C}], are plotted in Fig.
\ref{fB} for various values of $w$. One can see
that $B / \nu $ is $\xi$-dependent and takes greater values for
narrower solitons. On the other hand, the corrections in $C$
coming from the $\nu$-term are greater for wider solitons, and are
nearly $\xi$ independent.

In the previous equations we can make the change of variable $
\alpha \equiv \chi t + \theta $, after this change it results a
system of autonomous differential equations for $\alpha$ and $\xi$
\bea
\frac{d\alpha}{dt} &=& \chi + B(\xi) \\
\frac{d\xi}{dt} &=& C(\xi) \sin(2\alpha) \label{dxidt} \;.
\eea
The differential equation for the orbits is
\be
\frac{d\alpha}{d\xi} = \frac{\chi+B(\xi)}{C(\xi)\sin(2\alpha)}\;,
\ee
that gives the equation for the orbits $ \alpha = \alpha(\xi) $.
Deriving with respect to time Eq. (\ref{dxidt}) and substituting
the equation for the orbits, $ \alpha = \alpha(\xi) $, we obtain a
second order evolution equation for $\xi$ in terms of an effective
force, $ F_{eff} $, that only depends on $\xi$,
\bea
\frac{d^2\xi}{dt^2} &=& F_{eff}(\xi) \\
F_{eff}(\xi) &=& \frac{dC(\xi)}{d\xi}\, C(\xi)
\sin^2[2\alpha(\xi)] + 2 C(\xi) [\chi + B(\xi)] \cos[2\alpha(\xi)] \;.
\eea
This evolution equation (multiplying by $d\xi/dt$ and integrating
over time) leads to the energy conservation law
\bea
&& \frac12 \left(\frac{d\xi}{dt}\right)^2 + V_{eff}(\xi) =
constant\;,
\label{conlaw}\\
&& V_{eff}(\xi) = - \int_0^\xi \! d\xi' \, F_{eff}(\xi') \;,
\label{Veff}
\eea
where $V_{eff}$ is the effective potential. Thus, the soliton in
the adiabatic approximation can be seen as a particle moving in an
effective potential. This interpretation is very useful to
understand the Bloch oscillations and its freezing.

\section{Dynamics in the presence of the $\nu$-term} \label{sDynamics}

The evolution of soliton initial states of the form
\be
\psi_{n} = {\frac {\sinh(2\,w)}{\cosh[2\,w (n-\xi_0)]}}
\ee
is characterized by two main phenomena: the Bloch oscillations and
its freezing.
\subsection{Bloch oscillations}
\begin{figure}[h]
\[
\includegraphics[width=6cm]{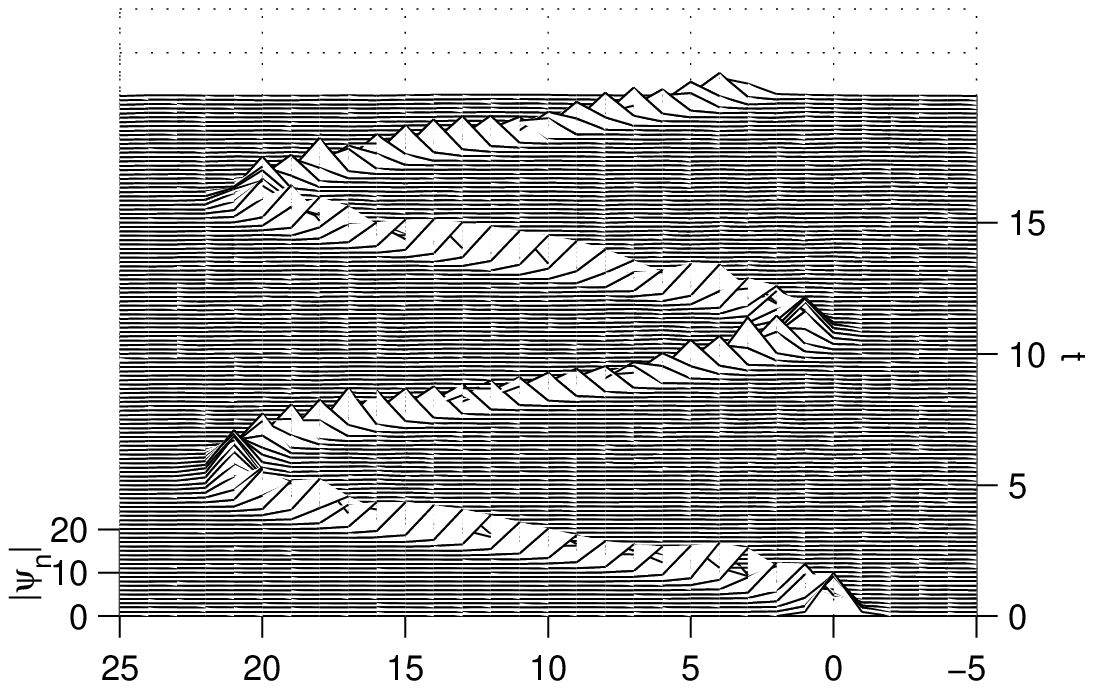}
\includegraphics[width=6cm]{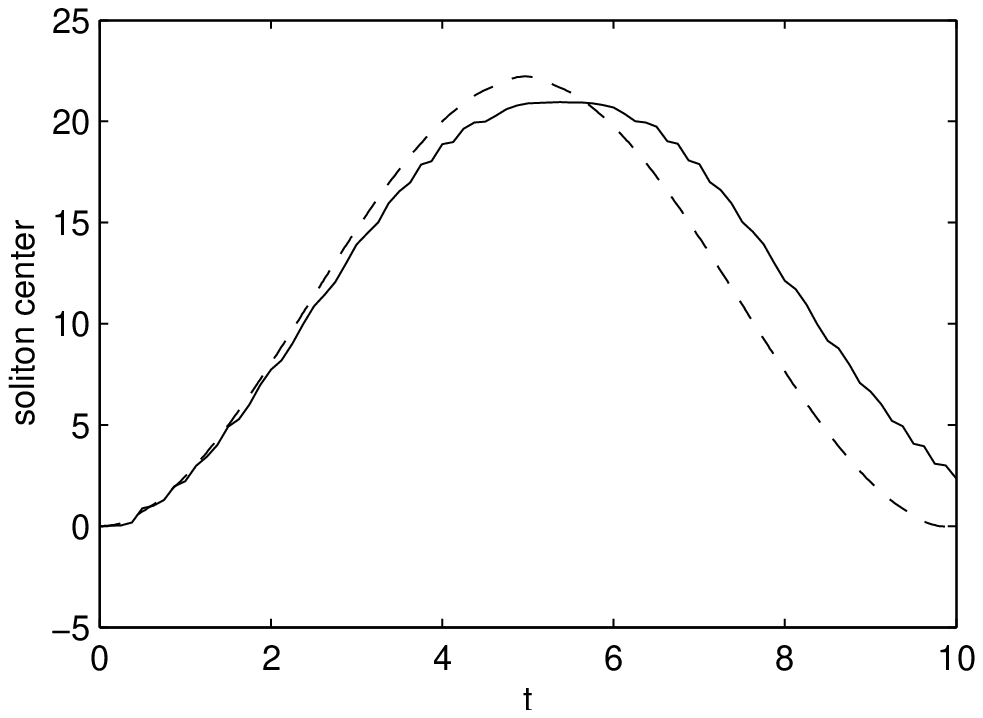}
\]
\caption{{\em (a)} Bloch oscillations of a soliton [full evolution,
Eq.~\eqref{DNLSeqf}] in a Schr\"odinger chain with $ \nu = -0.01 $.
{\em (b)} Dynamics
of the center of the soliton [adiabatic approximation,
Eqs.~\eqref{dtheta}-\eqref{dw}] for $ \nu =
-0.01 $ (full line) and for $ \nu = 0 $ (dashed line). Both
figures with initial conditions $ w_0 = 1.5 $, $ \xi_0 = 0 $ and $
\theta_0 = 0 $, in external field $ \chi = - 0.3 $.}
\label{fBloch}
\label{fXav}
\end{figure}

The evolution of these states present dynamical localization in
the form of Bloch oscillations, see Fig. \ref{fBloch} (as in the
$\nu = 0$ case \cite{Scharf}). This phenomenon is a consequence of
the discreteness of the system.

\begin{figure}[h]
\[
\includegraphics[width=6cm]{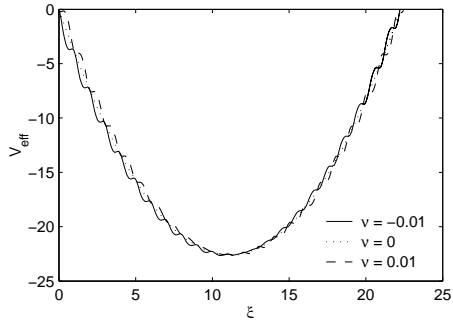}
\]
\caption{Effective adiabatic potential for a
soliton with initial conditions $ w_0 = 1.5 $, $ \xi_0 = 0 $ and $
\theta_0 = 0 $ in a Schr\"odinger chain with an
external field $ \chi = - 0.3 $ for various values of $ \nu $.}
\label{fVeff}
\end{figure}

The dynamical localization can be understood in the adiabatic
approximation (see Eq. \ref{Veff}). This approximation predicts
that the soliton center position, $\xi$, feels a trapping
effective potential (see Fig. \ref{fVeff}), and the turning points
of the motion are for $ 2(\chi t + \theta) $ multiple of $\pi$
(see Eq. \ref{dxir}). The adiabatic approximation gives good
results for the amplitude and the period of the Bloch
oscillations, and we also observe that the soliton width ($ 1/w $)
remains approximately constant as the adiabatic approximation predicts.

The $\nu$ term changes the period and amplitude of the Bloch
oscillations, and introduces new (fast) frequencies in the
dynamics [Fig.~\ref{fXav}]. These differences can be seen as a
consequence of the changes in the adiabatic effective potential,
see Fig. \ref{fVeff}. For $\nu>0$ the lattice sites have an
effective attractive interaction for the soliton center; thus the
effective potential displays local minima near the lattice sites.
Conversely for $\nu<0$ the effective interaction is repulsive, and
the local minima are located in the intersite regions.

The evolution of the center of the soliton is correctly predicted
by the adiabatic approximation (Eqs.~(\ref{dthetar})-(\ref{dwr}))
at early times. Later the effects of the radiative decay and the
change of shape of the soliton became important, and the
description of these effects falls beyond the scope of the
adiabatic approximation. These nonadiabatic effects become more
important for increasing $ |\nu| $.
 Another phenomenon that cannot be studied within the adiabatic
 approximation is the emergence of chaotic behavior \cite{chaos},
 the adiabatic approximation do not account of this effect
 because it reduces the computation of the evolution to the
 integration of the conservation law in Eq. (\ref{conlaw}).

\subsection{Freezing of the Bloch oscillations}

\begin{figure}[h]
\[
\includegraphics[width=6cm]{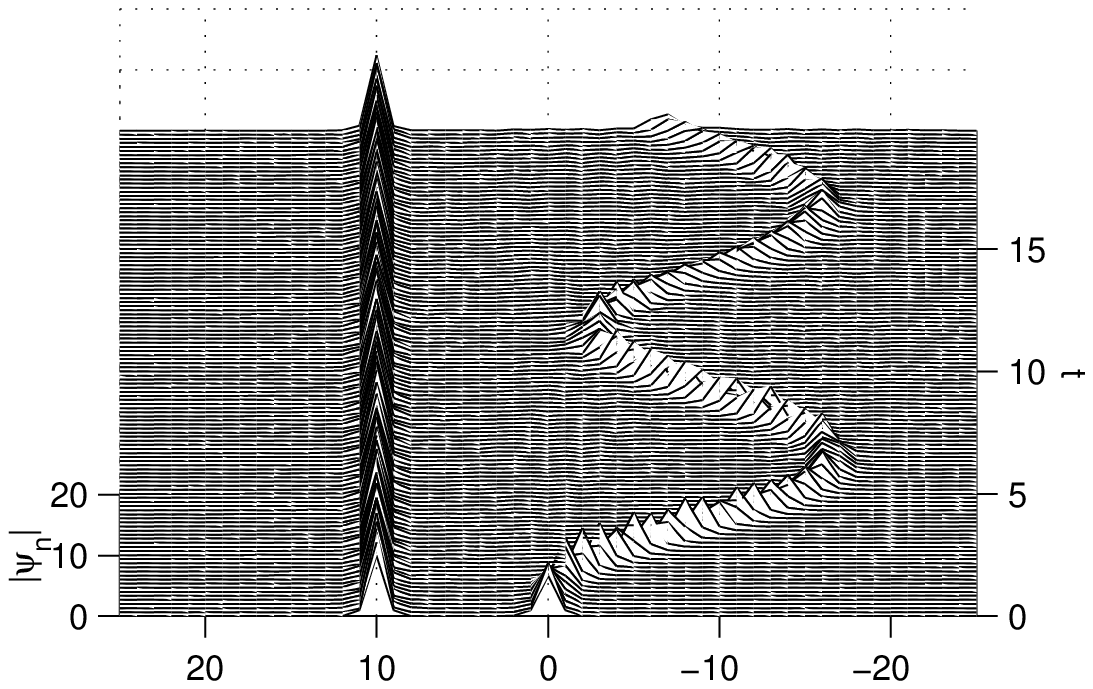}
\includegraphics[width=6cm]{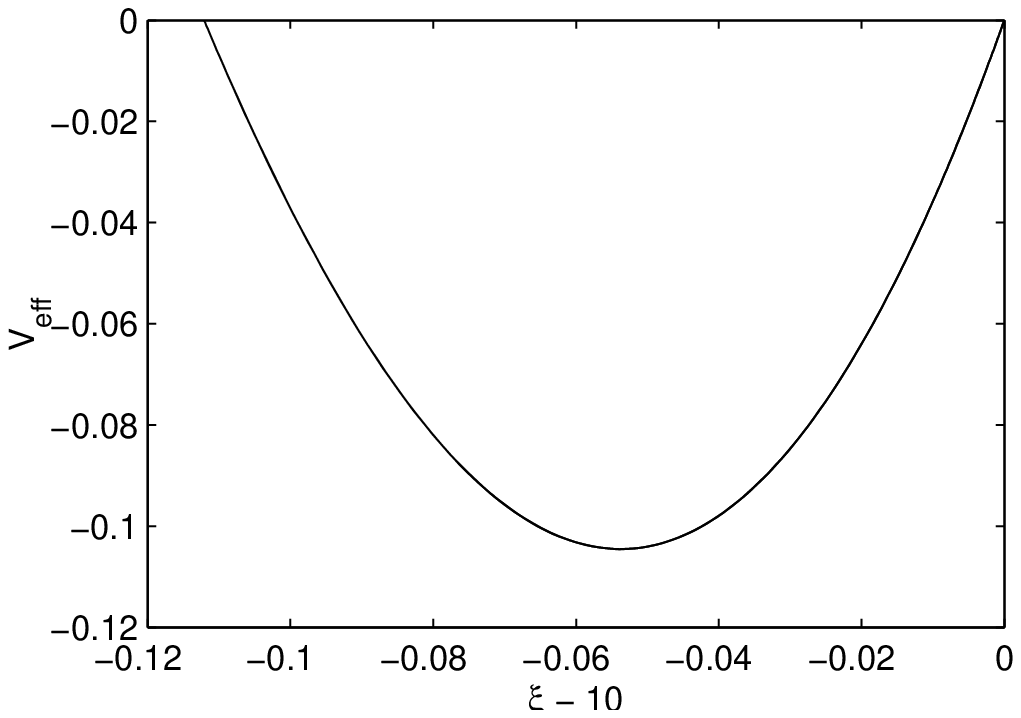}
\]
\caption{{\em (a)} Narrow soliton ($ w_0 = 1.5 $ and $ \xi_0 = 10
$); and wide soliton ($ w_0 = 1.3 $ and $ \xi_0 = 0 $) evolutions
in a Schr\"odinger chain with $ \nu = 0.02 $ in an external field
$ \chi = 0.3 $, both with $ \theta_0 = 0 $. {\em (b)} Effective
adiabatic potential for the narrow soliton.}
\label{fFreez}
\label{fVeffFreez}
\end{figure}

The main new phenomenon reported in this article is the freezing
of the Bloch oscillations due to an {\em homogeneous}
integrability breaking term (the $\nu$-term). This freezing is a
trapping of the soliton in a lattice site (or in a intersite
region), that happens in a region of the parameter space
[Fig.~\ref{fFreez}]. This is an {\em intrinsic localization}
effect due to characteristics of the lattice chain, not a
localization induced by lattice inhomogeneities \cite{KonoAdiabatic}.

The freezing of the Bloch oscillations can be easily explained in
the adiabatic approximation, where it corresponds to the center of
the soliton variable being trapped in a local minimum of the effective
potential. See Fig. \ref{fVeffFreez}.

\begin{figure}[h]
\[
\includegraphics[width=6cm]{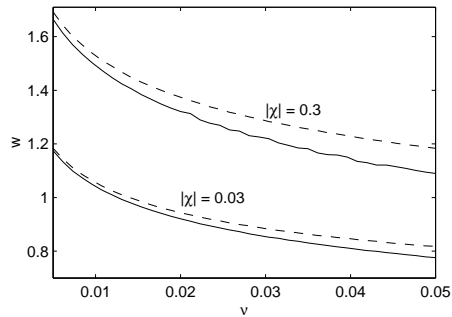}
\]
\caption{Bloch oscillations are frozen in the regions above
the curves for values of the external field smaller than those
indicated in the curves (taking for the other parameters $\theta_0
= 0 $ and $\xi_0 = 0$). Full lines: computation with the complete
equations of motion. Dashed lines: computation with the adiabatic
approximation.}
\label{fXcont}
\end{figure}

\begin{figure}[h]
\[
\includegraphics[width=6cm]{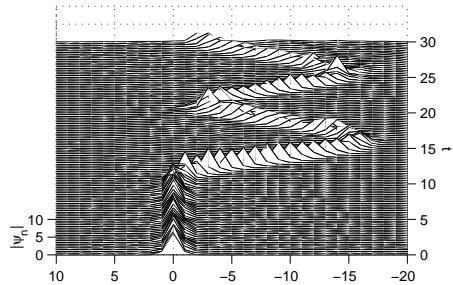}
\]
\caption{Dynamical defrosting due to nonadiabatic effects (radiation and
change of soliton shape) of a soliton with initial conditions
close to the frozen-unfrozen limit in parameter space. $ w_0 =
1.2633 $, $ \xi_0 = 0 $ and $\theta_0 = 0 $ in a Schr\"odinger
chain with an external field $ \chi = 0.3 $ for $ \nu = 0.025 $.}
\label{fUnfreez}
\end{figure}

Therefore, in the case $\nu>0$, the freezing emerges for solitons
centered near lattice sites (and with $d\xi/dt \simeq 0$); while
in the case $\nu<0$, it emerges when the soliton center is near
the middle of an intersite region. In both cases the phenomena is
favored for narrow solitons (greater $w$), strong $\nu$-terms
(greater $\nu$) and weaker external fields (smaller $|\chi|$). See
Fig. \ref{fXcont}.
The parameter region predicted by the adiabatic approximation
[Eqs.~\eqref{dtheta}-\eqref{dw}] is
in good agreement with the results of the numerical evolution of
the full equation of motion [Eq.~\eqref{DNLSeqf}].
However, the adiabatic approximation
is not able to describe the unfreezing at early times, that can
only be observed when the parameters are very close to the limit
of the freezing region [Fig.~\ref{fUnfreez}]. The reason is that
this unfreezing is related to non-adiabatic effects (change of the
soliton shape and radiation).

\section{Conclusions} \label{sConclusions}

These results show that inhomogeneity in the lattice is {\em not}
required to freeze the Bloch oscillations, {\em i.e.}, to trap the
soliton in a region with a size of the order of the intersite
distance in the lattice. We show that the {\em discrete
translational invariant} integrability-breaking term that makes
the transition between AL and DNLS models can freeze the nonlinear
Bloch oscillations, {\em i.e.}, gives rise to an {\em intrinsic
localization}. This phenomenon can be
understood as the trapping of the soliton center position in a minimum
of the effective potential that we obtain in the collective variable
or adiabatic approximation. The integrability-breaking terms considered, the
$\nu$-term, produces an effective attractive (for $\nu>0$) or
repulsive (for $\nu<0$) interaction with the lattice sites, that
gives rise to local minima in the effective potential where the
soliton can be trapped. This $\nu$-term has also the effect of
changing the main frequency of the Bloch oscillation and
introducing faster new frequencies in the dynamics, that can also be
understood as a consequence of the change in the effective potential.

These new phenomena show a richer dynamics of the DNLS equation,
increasing the interest and potential applications of this model,
and they also suggest the possibility of other new effects in the
presence of time varying external forces.

\section*{Acknowledgments}
We thank Vladimir Konotop and Angel S\'anchez for their comments.
We acknowledge support from the European Commission, Universidad
Complutense de Madrid (Spain) and MCyT (Spain) through grants
HPRN-CT-2000-00158, PR1/03-11595 and BFM2003-02547/FISI,
respectively.

\end{document}